\documentclass[12pt,a4paper]{article}
\usepackage{graphicx,psfrag,bbm,latexsym,color,dcolumn,bm,dsfont,bbm,color,mathrsfs,bbold,latexsym,amsfonts,amssymb}
\usepackage{psfrag}
\usepackage{amsmath}
\usepackage{amssymb}
\usepackage{amsfonts}

\topmargin= - 1.0cm

\newcommand{\mb}[1]{{\mathbf #1}}
\newcommand{\mc}[1]{{\mathcal  #1}}

\title{Large deviations and the Boltzmann entropy formula}

\author{Giovanni Jona-Lasinio\\
Dipartimento di Fisica, Universit\`a di Roma ``La Sapienza'', \\
and Istituto Nazionale di Fisica Nucleare,\\
Piazzale A. Moro 2, Roma 00185, Italy\\
E-mail: gianni.jona@roma1.infn.it}

\date{}
\begin{document}
\maketitle

\begin{abstract}
In the last decades the theory of large deviations has become a main tool
in statistical mechanics especially in the study of  non--equilibrium.
In a rational reconstruction of the story one must recognize the ideal
connection and debt of some recent work, to discussions taking place at the beginning
of the twentieth century. The famous equation $S=k\ln W$
usually attributed to Boltzmann, actually written in this final form 
by Planck on his route to the quantum hypothesis, was interpreted
by Einstein as a large deviation formula. This interpretation, on which he 
based his theory of thermodynamic equilibrium fluctuations, has been
a source of  inspiration in  recent developments of non--equilibrium statistical mechanics. In this paper we briefly illustrate this aspect.     
\end{abstract}

\section{The interpretation of the Boltzmann formula at the beginning of the XXth century}
It is typical in physics that the same formula or related 
equations may be interpreted differently by different authors
or in different times. Think for example of the Maxwell equations before and
after the special relativity: the concept of aether disappears. 
Or it may happen that an incomplete
formulation takes its final form in the work of an author not coinciding
with the first proponent who remains however the father of the idea.
A case of this type is to a certain extent the principle of inertia 
attributed to Galileo who however did not distinguish between a uniform 
motion in a straight line and a uniform circular motion. 
In fact his followers, in particular his pupil Torricelli,  gave the more restricted 
modern interpretation. The case we are considering  in this paper is
particularly interesting as three main figures of modern physics are involved.

\medskip

The fundamental relationship  between entropy and probability
\begin{equation}
\label{S}
S=k\ln W, 
\end{equation}
where $W$ is the so called number of {\sl complexions} or
thermodynamic probability, that is the number of microscopic states compatible with the values of the macroscopic parameters, and $k$ the Boltzmann
constant, was written in this form for the first time by Planck. 
It is in fact a fundamental step in his theory of the black body radiation leading
to the quantum hypothesis see e.g. \cite{pl0}. A detailed discussion is given
in his book {\sl The Theory of Heat Radiation} \cite{pl}.
This is a translation of the german second edition published by Planck
in 1913 and is based on lectures given in 1906-7. The formula was   
used by Einstein as a large deviation formula on which he based 
his theory of opalescence in fluids at equilibrium \cite{ei}. This is probably the first appearance of large deviation estimates in statistical physics. 
We now  briefly recall the Boltzmann formulation and the point of view of Planck and Einstein. 

\medskip

\medskip

{\sl Boltzmann}. The standard reference for Boltzmann is {\sl Lectures on Gas Theory} \cite{bo}.
After introducing in section $5$ the quantity $H$, often called Boltzmann entropy, 
\begin{equation} 
\label{H}
H=\int f\ln f d\omega,
\end{equation}
where $f$ is the distribution of molecules in the velocity space of a single molecule at time $t$ and $d\omega$ a volume element in this space, he proves that $H$, due to the effect of molecular collisions, decreases monotonically with
time, the famous $H$-theorem. In the subsequent section $6$ he discusses the 
probabilistic interpretation of $H$ for the ideal gas  where an explicit calculation is possible. The physical interpretation is analysed
in section $8$ where he remarks, by computing the entropy of the ideal gas, that except for a proportionality factor, the sign and an additive constant, it can be identified with $H$. The proportionality factor is the same for all gases and coincides with the gas constant $R$.

\medskip

\medskip

{\sl Planck}. Planck writes the connection between entropy and probability in the form \eqref{S} with no additive constant and this is the main difference with respect to Boltzmann. After providing a general argument showing that the relationship between entropy and probability must be of the form
\begin{equation} 
\label{gen}
S=k\ln W + const.,
\end{equation}
he emphasizes that $k$ must be a universal constant the same for a terrestrial as for a cosmic system. The constant is later identified with the so called Boltzmann constant $k=R/N$, $R$ is the gas constant, $N$ Avogadro's number,    and evaluated numerically. The argument goes as follows. Assume that there exists a general
relationship between entropy and probability $S=f(W)$ and consider a system made of two far apart subsystems so that $W=W_1W_2$. Since the entropy from thermodynamics is additive we must have $f(W_1W_2)=f(W_1)+f(W_2)$. The general solution of this functional equation is given by \eqref{gen}.     
Let us quote some of  his comments \cite{pl}. 

\medskip

{\sl ``Nevertheless our equation \eqref{S} differs in its meaning
from the corresponding one of Boltzmann. Firstly, Boltzmann's equation lacks the factor $k$, which is due to the fact
that Boltzmann always used gram-molecules, not the molecules themselves, 
in his calculations.
Secondly, and this is of greater consequence, Boltzmann
leaves an additive constant undetermined in the entropy $S$ as is done in the whole of classical thermodynamics, and accordingly there is a constant factor of
proportionality, which remains undetermined in the value of the probability
$W$. 

In contrast with this we assign a definite absolute value to the entropy $S$.
This is a step of fundamental importance which can be justified only by its consequences. As we shall see later, this step leads necessarily to the 'hypothesis of quanta' and moreover it also leads, as regards radiant heat, to a definite law of distribution of energy of black radiation, and, as regards heat energy of bodies, to Nerst's heat theorem.''   }

\medskip

There is in fact a direct connection between determining the additive  constant and the hypothesis of quanta. Fixing the constant makes the relationship between entropy  and probability well defined. 
The question is then how to calculate $W$. Specifying the state of  a molecule requires giving the coordinates and momenta corresponding to its degrees of freedom.
These parameters form a continuous space so that in order to count     
the number of microscopic states compatible with the macroscopic state we have to divide it into separate parts, the region elements in the terminology of Planck. Then we can count the number of ways
to  assign the molecules to the different region elements. The  hypothesis of quanta is that these region elements have a definite finite magnitude, the same for all of them.

\medskip

We quote again Planck

\medskip

{\sl `` .....the calculation of the entropy of a system of $N$ molecules in a given thermodynamic state is, in general, reduced to the single problem of
finding the magnitude  $G$ of the region elements of the state space. That such a definite finite quantity really exists is a characteristic feature of the theory we are developing, as contrasted to that due to Boltzmann, and forms the content of the so called hypothesis of quanta. ....... As is readily seen, this is an  immediate consequence of the proposition that the entropy $S$ has an absolute, not merely relative, value; ......''}

\medskip

\medskip

{\sl Einstein}. Einstein takes quite a radical view \cite{ei} on Boltzmann
formula 

\medskip

{\sl ``Boltzmann's principle can be expressed by the equation
\begin{equation}
\label{S_0}  
S={\frac RN}\ln W + const.,
\end{equation}}

{\sl W is commonly equated with the number of different possible ways
(complexions) in which the state considered - which is incompletely defined in the sense of a molecular theory, by parameters of a system - can conceivably be realized. To be able to calculate $W$, one needs a complete theory of the system
under consideration. If considered from a phenomenological point of view
equation \eqref{S_0} appears devoid of content. 

However, Boltzmann's principle does acquire some content independent of any
elementary theory if one assumes and generalizes from molecular kinetics the proposition that the irreversibility of physical processes is only apparent.''}

{\sl ``For let a state of a system be determined in the phenomenological sense by the
variables $\lambda_1....\lambda_k$ that are observable in principle. To each state $Z$ there corresponds a combination of values of these variables. If the system is externally closed then the energy - and in general no other function
of the [microscopic] variables - is constant. Let us think of all the states of the system that are compatible with the energy value, and denote them by $Z_1, ... ,Z_i$. If the irreversibility of the process is not one of principle, in the course of time, the system will pass through these states again and again. On this assumption one can speak of the probability of individual states in the following sense: suppose we observe the system for an immensely long time $\theta$ and deternine the fraction $\tau_1$ during which the system is in the state $Z_1$;
then $\tau_1/\theta$ represents the probability of the state $Z_1$. The same holds for the probability of the other states. ...........''}

\medskip

{\sl ``It follows from \eqref{S_0} that
\begin{equation}
W=const \cdot e^{{\frac NR}S}.
\end{equation}
The order of magnitude of the constant is determined by taking into account
that for the state of maximum entropy (entropy $S_0$) W is of the order  of magnitude one, so that we then have, with order of magnitude accuracy,
\begin{equation}
\label{pr}
W= e^{{\frac NR}(S-S_0)}.
\end{equation}
From this we can conclude that the probability $dW$ that the quantities
$\lambda_1....\lambda_k$ lie between $\lambda_1$ and $\lambda_1 + d\lambda_1$......$\lambda_k$ and $\lambda_k + d\lambda_k$ is given, in order of
magnitude, by the equation
\begin{equation}
\label{pr1}
dW= e^{{\frac NR}(S-S_0)} d\lambda_1...d\lambda_k,
\end{equation}
in the case when the system is determined only incompletely by $\lambda_1....\lambda_k$.''}

\medskip

The article \cite{ei} continues by expanding the entropy around the maximum $S_0$ and developing the
theory of opalescence of fluids as due to fluctuations of the thermodynamic state of the system. 

\medskip

\medskip

The attitudes of Planck and Einstein reflect a deep change in the character
of theoretical physics at the beginning of the XXth century, a change largely
due to them. While the interpretation of theories of the XIXth century was often  supported by mechanical
models (Maxwell) and/or philosophical credos (Helmholtz, Hertz), a principle
of modern theoretical physics is that a hypothesis has to be
evaluated by its verifiable empirical consequences, independently of any model possibly supporting it. This implies in particular much more freedom in the 
interpretation and use of formulas or equations. The only requirement is that
all symbols appearing in them represent quantities accessible, at least in principle, experimentally. It is in this logic that Planck disposes of the undetermined constant in Boltzmann formula or Einstein uses entropy, which is thermodynamically accessible, to calculate probabilities. Einstein's point of view on Boltzmann is further clarified in a lecture at the Z\"urich Physical Society \cite{ei1}.

\medskip

\medskip

In 1931 Onsager \cite{o}, in the same vein as \cite{ei} that he quoted, made use of Boltzmann formula in the study of fluctuations in non--equilibrium
phenomena  under the condition of small deviations  from equilibrium. The theory was developed further by Onsager and Machlup in \cite{om} where fluctuations of time trajectories of thermodynamic variables were considered under the same hypotheses.    
In the next section we discuss how it is possible to give an effective phenomenological meaning to
a formula like \eqref{pr} in the more general situation of stationary states
non necessarily close to equilibrium.
Typically we think of systems in contact with thermostats at different temperatures and/or
reservoirs characterized by different chemical potentials and under the action of external fields. The result  represents a step forward with respect to Onsager and Onsager--Machlup theory. 

\section{Non isolated systems out of equilibrium}
 The main idea of a recently developed study of non--equilibrium stationary states known under the name of Macroscopic Fluctuation Theory \cite{mft} is to start with the analysis of large macroscopic fluctuations in such states. This means
 searching for extensions of formulas like \eqref{pr} or \eqref{pr1}.
 The first difficulty encountered is how to define theoretically and empirically  non--equilibrium analogs of thermodynamic functionals like entropy or free energy. In other words the question is what to put in the exponent.  

\medskip

To  discuss this 
problem we need to analyze the meaning of the difference $S-S_0$ in    
\eqref{pr}. In an isolated system energy is conserved so that, if the volume remains constant,
$S-S_0 = - {\frac {F-F_0}{T}}$  where $F$ is the Helmholtz free energy. The quantity $F_0 - F$ represents the minimal work to bring the equilibrium state to the state corresponding
to $F$ at constant temperature and volume. 

\medskip

The concept of minimal work is meaningful also in non--equilibrium 
and can be taken as a generalization of the free energy. However we have to show that it can be calculated in terms of macroscopic quantities paralleling the 
calculation of the equilibrium entropy in terms e.g. of specific heats. As far as the probability of a state is concerned, also in nonequilibrium 
it can be given a phenomenological meaning  in terms of ergodic theory using Einstein argument.

\medskip

The difference with respect to an isolated system, or one in equilibrium with its environment, is that currents are flowing through the system. Currents are
empirically related to spacial gradients of the thermodynamic variables and to
external fields. The relationship is typically expressed by diffusion coefficients and conductivities and it is reasonable to expect that the minimal work to create a fluctuation can be calculated in terms of these quantities. In fact it can be shown that  
the calculation of the minimal work can be reduced to the solution of a variational problem if we  restrict sufficiently the class of systems considered. We shall concentrate on purely diffusive systems. 

\medskip

The macroscopic dynamics of diffusive systems is described by hydrodynamic
equations provided by conservation laws and constitutive equations, that is equations expressing the current in terms of the thermodynamic variables.
More precisely on the basis of a local equilibrium assumption, at the macroscopic level the system is completely described by a local multicomponent density $\rho(t,x)$ and the corresponding
local currents $j(t,x)$, see e.g. \cite{fi}. Their evolution is given by the
continuity equation and the constitutive equation which expresses the
current as a function of the density. Namely,
\begin{equation}
\label{2.1}
\begin{cases}
\partial_t \rho (t) + \nabla\cdot j (t) = 0,\\
j (t)= J(\rho(t)),
\end{cases}
\end{equation}
where we omit the explicit dependence on the space variable $x$.

\medskip

For diffusive systems the constitutive equation takes the form
\begin{equation}
\label{2.2}
J(\rho)  = - D(\rho) \nabla\rho + \chi(\rho) \, E,
\end{equation}
where the \emph{diffusion coefficient} $D(\rho)$ and the \emph{mobility}
$\chi(\rho)$ are $d\times d$ symmetric and positive definite matrices, $E$ is an external field. 
These equations must be supplemented with boundary conditions expressing
the interaction with the reservoirs.
The diffusive regime is revealed in \eqref{2.2} by the absence of  inertial terms. Equations \eqref{2.1} and \eqref{2.2}  are macroscopic
dynamical phenomenological laws of wide applicability.  The input
they require are the transport coefficients $D, \chi$ which are measurable
quantities. Systems for which the current at time $t$ can be expressed in terms of thermodynamic variables at the same time, as in \eqref{2.2}, are called Markovian independently
of whether the microscopic dynamics is Markovian or not \cite{ca}. 
 
\medskip

We now sketch how for these systems the calculation of the minimal work to create a fluctuation can  be  reduced
to the solution of a macroscopic variational principle, so that a detailed microscopic theory is not necessary. An important remark: the theory of stationary states includes as a particular case equilibrium and the usual thermodynamic
free energy  can be recovered via a dynamical calculation. This follows from the fact that in local equilibrium, which
is necessary for the validity of the phenomenological equations, there is
a relationship between the transport coefficients and the equilibrium free 
energy usually called Einstein relation. This is given by $D(\rho)=\chi(\rho)f''(\rho)$ where $f''(\rho)$ is the second derivative of the equilibrium free energy density. 

\medskip

Consider a system in a stationary state characterized by a time independent solution $\bar\rho$ of the hydrodynamic equations and suppose that, due to a fluctuation,  a value of the density $\rho_0$ is attained which is also the initial point of an arbitrary trajectory $\rho(t)$ .  
Let in addition $j(t)-J(\rho(t))$ be a fluctuation of the current with respect to the  value prescribed by the constitutive equation. The current $j$ and the density $\rho$ must always be connected by the continuity equation $\partial_t \rho (t) + \nabla\cdot j (t) = 0$. The cost of this fluctuation will consist of two terms: the cost necessary to create the initial condition and the cost necessary to follow the trajectory $(j(t),\rho(t))$. We shall denote the first term by $V(\rho_0)$ and in non--equilibrium will represent the analog of $S_0-S$. It turns out that the second term is proportional to the energy dissipated by the extra current  $j(t)-J(\rho(t))$
\begin{equation}
\label{r05}
\mathcal I_{[{T_0},{T_1}]}(\rho,  j)=\frac 14 \int_{{T_0}}^{{T_1}} \! \!dt \int_\Lambda
\! dx\, [j-J(\rho)] \cdot  \chi(\rho)^{-1} [j-J(\rho)].
\end{equation}
For a simple interpretation of this formula think of an electric circuit. In this case $\chi^{-1}$ is the resistance and the double integral in \eqref{r05} is the energy dissipated by $j(t)-J(\rho(t)$ according to Ohm's law. The
factor $1/4$ is fixed by the Gaussian nature of the stochasticity responsible for the fluctuations and by the consistency with equilibrium.
 
\medskip

Using the Markovian hypothesis we can now write the probability of the joint  fluctuations of density and current in a stationary state
\begin{equation}
\label{r21}
\begin{split}
& \mb P \big( (\rho_\epsilon (t), j_\epsilon (t))
\approx (\rho (t), j(t))\,,\, t\in [T_0, T_1] \, \big) \\
& \quad \asymp \exp\big\{ - \epsilon^{-d} \,
\mc R_{[{T_0},{T_1}]}(\rho,  j) \big\},
\end{split}
\end{equation}
where 
\begin{equation}
\label{r09}
\mc R_{[{T_0},{T_1}]}(\rho,  j) = V(\rho(T_0)) +
\mathcal I_{[{T_0},{T_1}]}(\rho,  j).
\end{equation}
Let us explain the meaning of the various symbols. The parameter $\epsilon$ is a dimensionless scaling factor,
i.e. the ratio between the microscopic length scale (typical intermolecular distance) and the macroscopic one.  The factor $\epsilon^{-d}$ is of the order of the number of particles in a
macroscopic volume. The role of  Avogadro's number in \eqref{pr} is played here
by $\epsilon^{-d}$. With  $\rho_\epsilon (t), j_\epsilon (t)$ we denote the empirical density and current corresponding to a coarse graining
over a small macroscopic volume. Clearly these quantities depend on $\epsilon$.

\medskip

It is not difficult to see that the functional $V(\rho)$ is related to
 $\mathcal I_{[{T_0},{T_1}]}(\rho,  j)$  by projection  
\begin{equation}
V(\rho)=
\inf_{ \substack{\rho(t), j(t) \, : \\ \nabla \cdot j = -\partial_t \rho\\
\rho(-\infty)=\bar\rho, \rho(0)=\rho}}
\mathcal I_{[-\infty,0]} (\rho , j),
\end{equation}
where $\bar\rho$ is the stationary solution.

\medskip

In a large deviation perspective equations \eqref{r21}-\eqref{r09} can be considered as a generalization of the Boltzmann entropy formula incorporating the macroscopic dynamics. They are applicable to equilibrium and non--equilibrium stationary states of diffusive systems. 
For the study of current fluctuations 
 Bodineau and Derrida proposed what they called an
{\sl additivity principle}  \cite{BD}. The predictions of this principle  coincide with those of \eqref{r21}-\eqref{r09}  when the most probable current fluctuations are
associated to time independent density profiles. However this is not always the case as it is revealed by the existence of dynamical phase transitions. The general fluctuation formulas \eqref{r21}-\eqref{r09} were established in \cite{letcurr}. They were inspired and supported by the study
of microscopic models, the stochastic lattice gases, but they can be taken as a
principle of the non--equilibrium thermodynamics of diffusive systems to be
validated by its consequences.

\medskip

\medskip

Equation \eqref{r21} has already been applied to several problems for which we refer to \cite{mft}. It is remarkable that when compared with microscopic models
amenable to an effective mathematical treatment its predictions coincide with
exact microscopic computations.

\medskip

\subsection*{Acknowledgments}
The content of this note is an outcome of innumerable discussions with L. Bertini, A. De Sole, D. Gabrielli and C. Landim in the course of our long--standing collaboration.  I am grateful to Kirone Mallick for a critical reading.

\end{document}